# Study on the Identification of Financial Risk Path Under the Digital Transformation of Enterprise Based on DEMATEL-ISM-MICMAC

Jie Dong[1]

[1] School of Statistics and Mathematics, Zhongnan University of Economics and Law, Wuhan, 430073, China
Author Email: 202021090069@stu.zuel.edu.cn

**Abstract**
Digital transformation challenges financial management while reducing costs and increasing efficiency for enterprises in various countries. Identifying the transmission paths of enterprise financial risks in the context of digital transformation is an urgent problem to be solved. This paper constructs a system of influencing factors of corporate financial risks in the new era through literature research. It proposes a path identification method of financial risks in the context of the digital transformation of enterprises based on DEMATEL-ISM-MICMAC. This paper explores the intrinsic association among the influencing factors of corporate financial risks, identifies the key influencing factors, sorts out the hierarchical structure of the influencing factor system, and analyses the dependency and driving relationships among the factors in this system. The results show that: (1) The political and economic environment being not optimistic will limit the enterprise's operating ability, thus directly leading to the change of the enterprise's asset and liability structure and working capital stock. (2) The enterprise's unreasonable talent training and incentive mechanism will limit the enterprise's technological innovation ability and cause a shortage of digitally literate financial talents, which eventually leads to the vulnerability of the enterprise's financial management. This study provides a theoretical reference for enterprises to develop risk management strategies and ideas for future academic research in digital finance.
**Keywords**: digital economy; enterprise financial risk; risk transmission path; DEMATEL; ISM-MICMAC

# 1 Introduction

With the new generation of information technologies such as big data, cloud computing, artificial intelligence, and blockchain as the driving force, digital transformation links the information of business, production, and management of enterprises. It calculates and feeds back effective details on strategic decisions for enterprise managers through powerful computing power, which ultimately empowers the whole process of the enterprise business value chain (Zhang et al., 2022). The development of digital technology provides opportunities for countries around the world to change existing business models, consumption patterns, socio-economic structures, related laws and policies, organisational models, cultural barriers, etc. (Berman, 2012). In the future, digital businesses will be 26% more profitable on average than non-digital businesses (Westerman et al., 2012). As the digital economy continues to unleash powerful dynamics and the new epidemic spreads worldwide, the trend will be for companies in all sectors to integrate their existing industries with digital technologies.

The proper use of information technology can effectively change the business structure and productivity of companies, enabling them to reduce costs and increase efficiency in the process of creating value (Correani et al., 2020). Digitalisation is a tool that helps companies avoid the difficulties of moving with the times since adopting new technologies is often accompanied by high learning costs and unknown risks (Al-Blooshi

& Nobanee, 2020). Companies also need to recognise that digital transformation, like most transformation practices throughout history, still has a variety of difficulties, with a particularly significant impact on financial management. Since financial risks usually have a relatively long latency, information technology can help companies achieve accurate ex-ante budgetary control. Companies tend to adopt digital transformation to manage and analyse financial data. It can reduce the prognostic risk of financial management and increase the sub-risks associated with information leakage and ownership disputes (Ionaşcu et al., 2022; Mosteanu & Faccia, 2020).

There have been studies focusing on the financial management risks of companies in digital transformation. (Zhao et al., 2022) found that the emergence of digital technology has overturned the traditional path of financial risk identification, and companies can rely on data mining and analysis to anticipate relevant risks holistically. Gonçalves et al. (2022) studied the perceptions of practitioners in corporate accounting departments about digital transformation and the role played by financial managers in digital transformation through a field interview method. Damerji & Salimi (2021) found through a case study that enterprises with the support of technologies such as cloud computing and artificial intelligence can achieve access to distributed ledgers and big data to automate decision-making at scale. However, the information quality risk to the enterprise increases. Teichert (2019) found through an extensive literature study that the profound changes generated by business models during the digital transformation of companies may sometimes create obstacles to their previous financial management systems.

Although previous studies in the literature have made a considerable contribution to the analysis of financial management risks in the context of corporate digital transformation, the theoretical research of corporate digital transformation in the academic community is still in its infancy (Kraus et al., 2021). Most of the studies focus on the impact on productivity, but there is a lack of systematic research results on the overall effect on the organisational structure of enterprises and the transmission mechanism of financial management risks after the transformation. Most studies are limited to machine learning and single-factor analysis, and there are still some gaps in the research on the interrelationship, influence logic, and factor classification of the risk factor system. Based on this, this paper proposes a path identification method based on DEMATEL-ISM-MICMAC for financial management risks under the digital transformation of enterprises. The technique aims to analyse the inner connection and hierarchical structure among the influencing factors of enterprise financial management risk in the context of digital transformation, analyse the dependency and driving force of factors, identify key influencing factors, and provide theoretical reference for enterprises to develop risk management strategies.

The hybrid approach constructed in this paper includes three parts, DEMATEL, ISM, and MICMAC. The DEMATEL model is a method for screening the main influencing elements in complex information systems, which can simplify the process of system structure analysis and analyse the correlation between factors in the system (Du & Li, 2021). The ISM model determines the structural hierarchy of system factors by constructing the adjacency relationships among them. Based on this method, the MICMAC model can explore the degree of dependency and driving effects among the influencing factors and the importance of each factor to the whole system (Mani et al., 2016). The contributions of this paper are as follows: (1) Present a complete list of factors influencing financial management risks in the context of the digital transformation of enterprises. (2) The financial risk and its transmission mechanism during the digital transformation of enterprises is an economic problem with a complex and large system of influencing factors. This paper tries to study it based on system engineering theory, using a coupled

DEMATEL-ISM-MICMAC approach that combines qualitative analysis and quantitative calculation. (3) Confirm the applicability of the systems engineering approach to socio-economic problems and provide a practical methodological library for the changes in the overall structure and risk transmission mechanisms of internal financial management of enterprises in the digital context.

The paper's discussion is structured as follows. Section 2 reviews the research literature on the digital transformation of enterprises and enterprise management risk. Section 3 presents the research methodology used in this paper and its computational process. Section 4 gives the model results, followed by a detailed explanation and analysis in Section 5. Section 6 summarises the research results of this paper and suggests future research directions.

# 2 Literature Review

## 2.1 Enterprise digital transformation

The current academic debate on the concept of digital transformation is still inconclusive. According to Zeng et al. (2021), digital transformation, starting with digital technology infrastructure, can lead to profound changes in the skills of employees, the way companies are organised and managed, and industrial business models and value chains, which can have either positive or negative impacts on companies. Goerzig & Bauernhansl (2018) argue that digital transformation is a fundamental change based on the self-reengineering of companies in the context of rapid changes in the global economy and market environment, which can largely change the relationship between companies and their suppliers, customers, and employees, thus helping them to improve productivity. Tabrizi et al.( 2019), on the other hand, argue that the discussion of digital transformation of companies cannot focus only on information technology; the process of digital transformation is indeterminable, and the process exposes flaws in the business strategy, internal organisational structure, brand culture and marketing strategy of companies.

## 2.2 Financial management risk

Financial risk includes broad risk and narrow risk. Broad risk is the possibility that the financial position deviates from expectations, mainly financing, investment, and profit distribution risks. In contrast, narrow risk refers to the possibility that the firm's cash flow is insufficient to pay creditors and meet other financial obligations (Cao & Zen, 2005). The Alexander Bathory model has been used to quantitatively measure financial risk, covering five financial aspects: capital structure, profitability, and capital ratio (Fu et al., 2012). The Bathory model is used to quantify the financial risk of a firm. Regarding the factors influencing financial risk, Chollet & Sandwidi (2018) argue that higher social responsibility performance of firms can significantly reduce the financial management risk of firms. Mahtani & Garg (2018) analysed the key factors that make a firm financially distressed regarding its primary business, government economic policies, profitability, leverage, market, and external factors. Belás et al. (2018), on the other hand, argue that factors such as entrepreneurial talent, corporate credit, and cash flow situation can have an impact on the financial management of the firm. By combing and summarising the literature and considering the actual situation, this paper obtains the structure chart of financial risk influencing factors, as shown in Fig. 1.

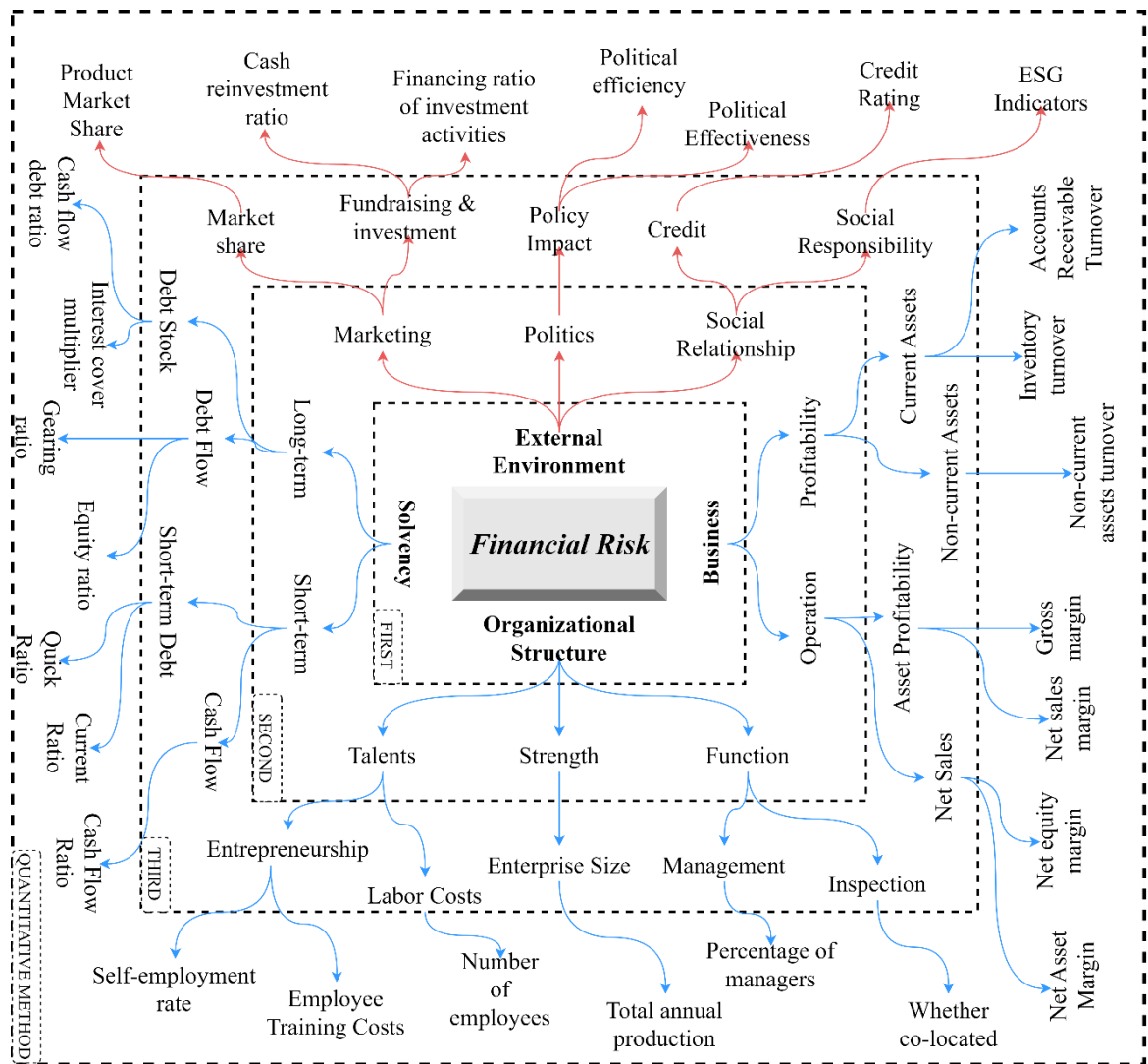

Figure 1 Structure of the factors influencing financial risk

## 2.3 Financial risk factors under enterprise digital transformation

Scholarly research on financial management risks in corporate digital transformation has been largely analysed in four aspects: external corporate environment, investment and financing performance, marketing capabilities, and organisational management. Vovchenko et al.(2019) discuss the risks resulting from the development of companies in an uncertain business environment in the digital context and the related need for new regulations, finding that inter-institutional cooperation support of companies is a necessary element of development. Suling et al.(2021) found that the role of fintech in suppressing corporate financial risk can not be achieved without effective financial regulation and that fintech development can also effectively alleviate the problem of corporate financing constraints. Hacioglu & Aksoy (2021) analyse the impact of technological developments and innovations on financial accounting, outlining the critical role of classical approaches to corporate financial management in firm performance and describing how emerging technological innovations will shape new approaches to financial management. (Dai & Lu, 2023) find that digital finance reduces corporate financial risk by alleviating corporate financing constraints and reducing inefficient investment and that this effect is more pronounced among SMEs and firms whose managers are not overconfident. Luo et al.(2021) analysed the impact of digital finance capabilities on corporate management, business innovation, and financial performance and found that the capabilities can affect corporate financial performance through sales, lending, and investment channels. Morgan et al.(2019) discuss the importance of individual digital financial literacy in the digital age and explore strategies and programs to promote digital financial education. Leeflang et al.(2014) found that filling talent gaps, adjusting organisational design, and implementing actionable metrics were how companies in various industries responded to the challenges of digital transformation through the interview method.

Uncertainty in the market environment and uncontrollable government policies are the biggest concerns that inhibit companies from taking the plunge into digital transformation. On the other hand, fintech is inherently beneficial to business growth and profitability. Information technology is not adequately regulated, and cooperation between companies is disagreed. In that case, society cannot effectively join forces to address the common problems faced in digital development. However, the talent management system, which is the source of support for business development, is easily ignored by decision-makers. On the one hand, if the top management lacks digital literacy, the overall transformation direction of the enterprise will be blind or even off; on the other hand, if the enterprise staff lacks digital literacy, the enterprise will reduce productivity due to the digital operation errors accumulated in the invisible. Based on literature collection and expert discussion, this paper finalises the indicators and their expressions as shown in **Tab. 1**.

Table1 Financial management risk indicator system under enterprise digital transformation

| Secondary indicators | Tertiary indicators | Indicator description | Indicator code | REF. |
|---|---|---|---|---|
| **External Environment** | Unreasonable economic policy guidance | The government does not introduce relevant policies or implement measures to suppress the market. | $x_1$ | (Hacioglu&Aksoy,2021), (Vovchenko et al., 2019) |
| | Slow implementation of government subsidies and incentives | Government plans to invest in stimulating business transformation but fails to act | $x_2$ | (Hacioglu&Aksoy,2021) |
| | Regulations related to digital development are unclear | Incomplete laws and rules on how to transform enterprises | $x_3$ | (Hacioglu&Aksoy,2021), (Vovchenko et al., 2019), (Suling et al., 2021) |
| | The business is located in an underdeveloped area. | The business is located in an area with a low frequency of economic activity. | $x_4$ | (Yuhui&Zhang,2023), (Suling et al.,2021) |
| | The market management in the industry is chaotic. | Market regulation is chaotic and prone to fluctuations in market prices and interest rates. | $x_5$ | (Vovchenko et al., 2019), (Suling et al., 2021) |
| | Overall digitalisation rate of the market is low. | Enterprise transformation is fast (slow), while overall market transformation is slow (fast) | $x_6$ | (Vovchenko et al., 2019) |
| **Assets and Liabilities** | Unreasonable structure of assets and liabilities | The debt portfolio is unreasonable, and there are significant debt repayment risks. | $x_7$ | (Hacioglu&Aksoy,2021), (Luo et al., 2021) |
| | The cash flow situation is not optimistic | Insufficient liquid assets reserve and low emergency capacity | $x_8$ | (Hacioglu&Aksoy,2021), (Luo et al., 2021) |
| | Small business scale | Total assets and the number of employees are small. | $x_9$ | (Yuhui&Zhang, 2023), (Hacioglu&Aksoy,2021), (Luo et al., 2021) |
| | There is a product inventory that is slow to sell | Excess product inventory, not | $x_{10}$ | (Hacioglu&Aksoy,2021), (Luo et al., 2021), (Leeflang et al. 2014) |
| | Digital transformation costs are too high. | reaching expected sales Digital products cost more or consume more manpower and material resources. | $x_{11}$ | (Leeflang et al. 2014) |
| | The investment planning plan is unreasonable. | Mistakes in investment asset allocation and investment object decision-making | $x_{12}$ | (Yuhui & Zhang, 2023), (Hacioglu&Aksoy,2021), (Luo et al., 2021) |
| | Forecasting Mistakes When Quantifying Investing | Use of inappropriate AI algorithms for market forecasting | $x_{13}$ | (Hacioglu&Aksoy,2021) |
| **Management Capacity** | Low level of digitalisation of core business | The main business still adopts the traditional management and operation mode. | $x_{14}$ | (Suling et al., 2021), (Luo et al., 2021) |
| | Insufficient technological innovation ability of enterprises | Lack of major original innovations and key core technologies | $x_{15}$ | (Morgan et al., 2019), (Suling et al., 2021) |
| | Blind pursuit of advanced digital products | Buy digital products with excess functionality. | $x_{16}$ | (Leeflang et al. 2014) |
| | Product market share reduction | A decrease in the total market demand for the products produced by the firm | $x_{17}$ | (Hacioglu&Aksoy,2021), (Leeflang et al. 2014) |
| | Inefficient cooperation with upstream and downstream enterprises | The level of management in the supply chain is low, and the cooperation between enterprises is not smooth | $x_{18}$ | (Vovchenko et al., 2019), (Suling et al., 2021) |
| | Low credit leads to financing difficulties. | Insufficient solvency, difficulty in obtaining financing from banks and institutions | $x_{19}$ | (Vovchenko et al., 2019), (Suling et al., 2021) |
| **Organisational Structure** | The enterprise transformation plan is not clear | Enterprise management has no plan for digital transformation | $x_{20}$ | (Yuhui&Zhang, 2023), (Luo et al., 2021) (Leeflang et al. 2014) |
| | Internal information management system confusion | The handover of employees from different departments and old and new employees is not smooth. | $x_{21}$ | (Luo et al., 2021), (Leeflang et al. 2014) |
| | Talent training and incentive mechanism is not perfect | Staff training and inspection mechanism is not perfect | $x_{22}$ | (Morgan et al.,2019), (Leeflang et al.2014) |
| | Lack of digitally literate talent | Insufficient digital capabilities of enterprise employees or high cost of digital talents | $x_{23}$ | (Morgan et al., 2019), (Hacioglu&Aksoy,2021), (Leeflang et al. 2014) |
| | Financial personnel make mistakes in operating digital products | Failure to keep financial accounts properly due to digital ignorance | $x_{24}$ | (Morgan et al., 2019), (Hacioglu&Aksoy,2021) |
| | The accounting department is co-located with other departments | The division of powers and responsibilities between the financial department and other stakeholders is blurred | $x_{25}$ | (Luo et al., 2021) |
| | Financial data leaked, lost, or falsified. | Damage to financial information due to factors such as imperfect information systems | $x_{26}$ | (Morgan et al., 2019), (Hacioglu&Aksoy,2021) |

# 3 Methodology

The financial management system in the context of enterprise digital transformation is complex, and to clarify the interactions among the influencing factors, this paper draws on three factor analysis methods of DEMATEL, ISM, and MICMAC in the discipline of system engineering to construct a causal system of financial management risk in the context of enterprise digital transformation and identify the key influencing factors in this system. The schematic flow of the methodology is illustrated in **Fig. 2**.

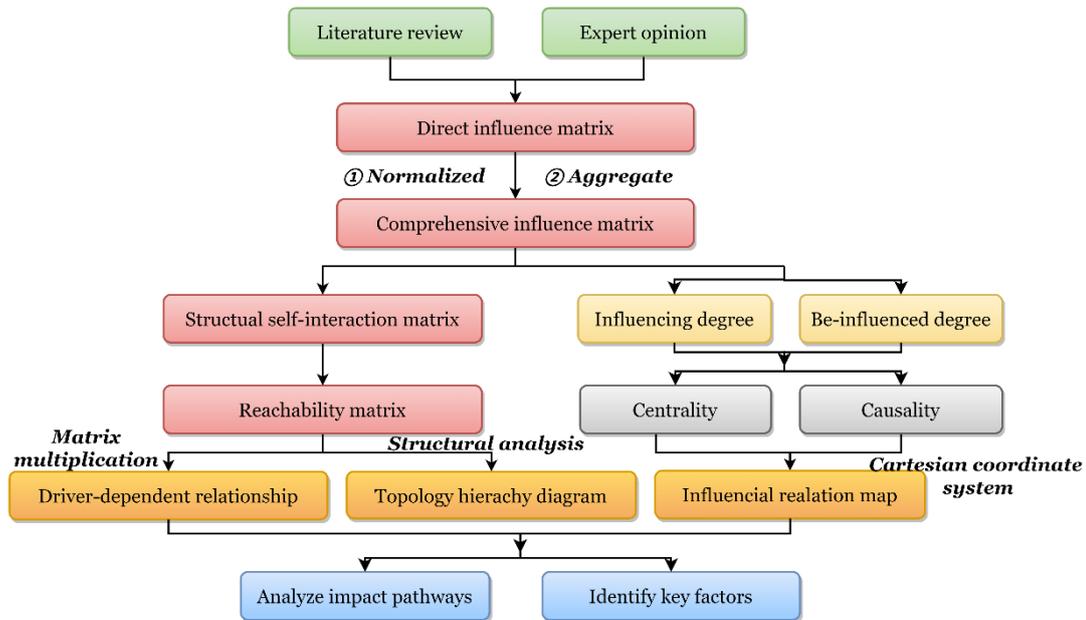

Figure 2 Flow chart of the methodology used in the present study

## 3.1 DEMATEL Method

Decision-making Trial and Evaluation Laboratory (DEMATEL) is a system analysis method proposed by American scholars in the early 1970s that uses graphical and matrix tools to explain problems, and it is very effective in dealing with complex social issues, especially those with ambiguous relationships between system elements (Wu & Lee, 2007). The model calculates the degree of influence of each element on other elements and the degree of influence through the logical relationship between the elements in the system and the direct influence matrix. It is used to calculate each element's degree of cause and centrality as the basis for constructing the model and thus determine the causal relationship between the elements and the position of each element in the system.

**Step 1**: Determine the system influence factors $x_1, x_2, \ldots, x_n$, and analyse the influence relationship between the factors, determine the degree of direct influence between the factors by expert scoring and other methods, and obtain the direct influence matrix of system $A$.

$$A = \begin{bmatrix} 0 & a_{12} & \cdots & a_{1n} \\ a_{21} & 0 & \cdots & a_{2n} \\ \vdots & \vdots & \ddots & \vdots \\ a_{n1} & a_{n2} & \cdots & 0 \end{bmatrix} = (a_{ij})_{n \times n} \qquad (1)$$

$a_{ij}(i = 1,2,\ldots,n\ ;\ j = 1,2,\ldots,n)$ denotes the degree of direct influence of the $i^{th}$ factor on the $j^{th}$ factor.

**Step 2:** Normalise the direct impact matrix $A$. The normalised direct impact matrix $G$ is usually obtained using the row and maximum method.

$$G = \frac{A}{\max_{1 \leq i \leq n} \sum_{j=1}^{n} a_{ij}} = (g_{ij})_{n \times n} \quad (2)$$

**Step 3:** Calculate the combined influence matrix $T$ for each influence factor in the system.

$$T = G(I - G)^{-1} = (t_{ij})_{n \times n} \quad (3)$$

It is an approximate calculation when the sample size $n$ is sufficiently large. $I$ is the unit matrix (Chen, 2021).

**Step 4:** Calculate the degree of influence and the degree of influence for each factor. Influence represents the combined influence of the factor $xi$ on other factors and the degree of being influenced $xi$ by all other factors. Influence $Fi$ and the degree of being influenced $Ei$ are shown in Eq. (4)-(5).

$$F_i = \sum_{j=1}^{n} t_{ij} \quad (4)$$

$$E_i = \sum_{j=1}^{n} t_{ji} \quad (5)$$

**Step 5:** Calculate the centrality and causality of each factor in the system. Centrality reflects the degree of importance of the elements $x_i$ in the system. Causality indicates the influence of $x_i$ on other factors, and a value greater than 0 means $x_i$ is the causal factor, while the opposite is the effect factor. Centrality $M_i$ and causality $N_i$ are shown in Eq. (6)-(7).

$$M_i = F_i + E_i \quad (6)$$

$$N_i = F_i - E_i \quad (7)$$

**Step 6:** Using the centrality and causality of the factors as a Cartesian coordinate system, we derive the cause-effect diagram of the factors in the system, analyse the importance of each factor and the cause-effect logical relationship between the factors, and make reasonable inferences and decisions for the complex economic and social problems under study.

## 3.2 ISM Method

Interpretative Structural Modeling (ISM) is a system engineering research method founded by American scholars in 1973, whose role is to study the structural relationships of the system and to analyse the rationality of the selection of system elements, giving the simplest hierarchical topology without loss of system functionality (Ahmad & Qahmash, 2021). The method applies the principle of the correlation matrix in graph theory, decomposes the complex system into complex subsystems according to certain methods, sorts out the logical structural relationships among them to find out the hierarchical level of each element, and finally forms a multi-level recursive structural model. The main calculation steps are as follows:

**Step 1:** Collect and organise the components of the system, and analyse whether there is an influence relationship among the factors using methods such as expert group discussion or the Delphi method to determine the influencing factors in the system $x_1, x_2, \ldots, x_n$.

**Step 2:** Determine whether there is a direct influence relationship between each element, and this relationship is mathematically expressed by the adjacency matrix $A = (a_{ij})_{n \times n}$. The definition of $a_{ij}$ is shown in Eq. (8).

$$a_{ij} = \begin{cases} 1, i \neq j \text{ when element } x_i \text{ has a direct influence relationship with } x_j \\ 0, i = j \text{ when the element } x_i \text{ has no direct influence on } x_j \end{cases} \quad (8)$$

**Step 3:** The reachable matrix $M$. The number 1 in the reachable matrix indicates that there is an influence path between element $x_i$ and other elements, while the number 0 indicates that there is no influence path between $x_i$ and another element. The calculation formula is shown in Eq. (9).

$$M = (A+I)^{k+1} \quad if \ (A+I) \neq (A+I)^2 \neq \cdots \neq (A+I)^k = (A+I)^{k+1} \ (k < n-1) \quad (9)$$

$I$ is the unit matrix, and the multiplication of the matrix is based on the Boolean algebraic algorithm.

**Step 4:** To divide the factors into different regions and levels, you need to obtain the reachable set $R$ and the prior set $Q$ and their intersection $N$. The reachable set represents the set of elements that can be reached from $x_i$ The reachable set is the set of factors that can be reached from the starting point, and the prior set is the set of factors that can be reached from $x_i$, and the intersection set $N = R \cap Q$ denotes the intersection of the reachable set and the prior set. The calculation principle is shown in Eq. (10)-(11).

$$R(x_i) = \{x_i | M_{ij} = 1\} \quad (10)$$

$$Q(x_i) = \{x_i | M_{ji} = 1\} \quad (11)$$

**Step 5:** Build the structural model based on the reachability matrix, including region division, inter-level division, and building the recursive explanatory model.

**Step 6:** Get the hierarchical distribution of each element. Based on the $R(x_i) \cap N(x_i) = R(x_i)$, determine the highest level of the set of elements $L_1$, and then remove these elements and determine $L_2$ until the bottommost element set $L_m$ is found.

### 3.3 MICMAC Method

Matrix-based Multiplication Applied to a Classification (MICMAC) is a scientific method used to analyse the influence and dependency relationships between factors in a system. The advantage of MICMAC is that it represents the interactions between influencing factors using an image of a drive-dependence matrix. MICMAC mainly applies the principle of matrix multiplication, and divides each element into autonomous, dependent, linkage, and driving factor families according to the nature of their influence. The basic idea of this method: if factor A directly influences factor b and factor b influences factor c, then there is an indirect influence relationship between factor a and factor c (Sindhwani & Malhotra, 2017). The basic analysis steps are as follows:

**Step 1:** Analyse and identify the influencing factors in the system $x_1, x_2, \ldots, x_n$. To determine the direct influence relationship existing between different elements, and then calculate the reachable matrix $M = (m_{ij})_{n \times n}$. The calculation method is the same as Eq. (9).

**Step 2:** Calculate the elements based on the reachability matrix$xi$ based on the reachability matrix. The greater the driving force, the greater the influence of the element$xi$ has a greater impact on other elements, and a greater dependency indicates $xi$ the greater the dependence on other influencing elements. The formula is as follows (12)-(13).

Calculate the driving power and dependence power of elements $x_i$ based on the reachability matrix. The greater the driving force, the greater the influence of the element $x_i$ on other factors, and the greater the dependence, the greater the dependency of $x_i$ on other influencing factors. The formula is as Eq. (12)-(13).

$$D_i = \sum_{j=1}^{n} m_{ij} \quad (12)$$

$$R_i = \sum_{j=1}^{n} m_{ji} \quad (13)$$

# 4 Results

## 4.1 Data Collection

According to the DEMATEL-ISM-MICMAC hybrid model introduced in the previous section, this paper is used the expert opinion method to collect data and invited eight experts in related fields (including 2 current PhDs, 3 university faculty members, and 3 senior corporate financial management practitioners) to score the mutual influence of each indicator, where a higher score indicates a greater degree of influence.

## 4.2 DEMATEL Result

To analyse the correlation and importance of the factors influencing corporate financial risk in the digital context based on the DEMATEL method, the data collected from eight questionnaires are averaged to obtain the direct influence matrix of corporate financial management risk in the digital context, and then calculate the degree of influence, being influenced, centrality and causality of the factors affecting risk. The results are shown in **Tab. 2**.

Table 2 Results of DEMATEL calculation

| Code | Indicator name | Influence | Influenced | Centrality | Causality |
|---|---|---|---|---|---|
| $x_1$ | Unreasonable economic policy guidance | 0.213 | 0.000 | 0.213 | 0.213 |
| $x_2$ | Slow implementation of government subsidies and incentives | 0.500 | 0.000 | 0.500 | 0.500 |
| $x_3$ | Regulations related to digital development are unclear | 0.426 | 0.000 | 0.426 | 0.426 |
| $x_4$ | The business is located in an underdeveloped area | 0.639 | 0.000 | 0.639 | 0.639 |
| $x_5$ | The market management in the industry is chaotic | 0.639 | 0.000 | 0.639 | 0.639 |
| $x_6$ | overall digitalisation rate of the market is low | 0.703 | 1.125 | 1.828 | -0.422 |
| $x_7$ | Unreasonable structure of assets and liabilities | 0.000 | 0.375 | 0.375 | -0.375 |
| $x_8$ | The cash flow situation is not optimistic | 0.000 | 2.591 | 2.591 | -2.591 |
| $x_9$ | Small business scale | 0.832 | 0.000 | 0.832 | 0.832 |
| $x_{10}$ | There is a product inventory that is slow to sell | 0.750 | 0.898 | 1.648 | -0.148 |
| $x_{11}$ | Digital transformation costs are too high | 0.000 | 0.980 | 0.980 | -0.980 |
| $x_{12}$ | The investment planning plan is unreasonable | 0.375 | 0.000 | 0.375 | 0.375 |
| $x_{13}$ | Forecasting Mistakes When Quantifying Investing | 0.125 | 0.375 | 0.500 | -0.250 |
| $x_{14}$ | Low level of digitalisation of core business | 0.469 | 0.281 | 0.750 | 0.188 |
| $x_{15}$ | Insufficient technological innovation ability of enterprises | 0.680 | 0.125 | 0.805 | 0.555 |
| $x_{16}$ | Blind pursuit of advanced digital products | 0.250 | 0.281 | 0.531 | -0.031 |
| $x_{17}$ | Product market share reduction | 0.375 | 0.320 | 0.695 | 0.055 |
| $x_{18}$ | Inefficient cooperation with upstream and downstream enterprises | 0.875 | 0.797 | 1.672 | 0.078 |
| $x_{19}$ | Low credit leads to financing difficulties | 0.500 | 0.000 | 0.500 | 0.500 |
| $x_{20}$ | The enterprise transformation plan is not clear | 0.422 | 0.000 | 0.422 | 0.422 |
| $x_{21}$ | Internal information management system confusion | 0.500 | 0.000 | 0.500 | 0.500 |
| $x_{22}$ | Talent training and incentive mechanism is not perfect | 2.151 | 0.000 | 2.151 | 2.151 |
| $x_{23}$ | Lack of digitally literate talent | 1.219 | 1.250 | 2.469 | -0.031 |
| $x_{24}$ | Financial personnel make mistakes in operating digital products | 0.625 | 1.688 | 2.313 | -1.063 |
| $x_{25}$ | The accounting department is co-located with other departments | 1.000 | 0.000 | 1.000 | 1.000 |
| $x_{26}$ | Financial data leaked, lost, or falsified | 0.000 | 3.180 | 3.180 | -3.180 |

Based on the results, an influence-influenced diagram and a centrality-cause diagram are drawn, as shown in **Fig. 3** and **Fig. 4**. The greater the degree of influence indicates the greater the effect of the factor on other factors, while the degree of being influenced reflects the extent to which the factor is disturbed by other factors. Similar to the degree of influence, the greater the degree of cause, the greater the influence of a factor on other factors, which can be called the cause element; conversely, it is called the result element. The greater the value of centrality, the stronger the correlation between the factor and other factors, i.e., the greater the role played by the factor in the system. In the centrality-causality diagram, the higher the centrality of a factor, the closer it is to the right side of the graph. The greater the centrality, the closer it is to the top of the chart.

In terms of the relevance of risk-influencing factors, the top 5 factors in terms of centrality are financial data leakage, loss or falsification ($x_{26}$), poor cash flow situation ($x_8$), lack of digitally literate people ($x_{23}$), financial staff's mismanagement of digital products ($x_{24}$), inadequate talent training and incentive mechanisms ($x_{22}$), which have a greater impact on the financial risk of enterprises in the digital context. Imperfect talent training and incentive mechanisms ($x_{22}$) have the highest impact, financial data leakage, loss or falsification ($x_{26}$), and lack of digitally literate talents ($x_{23}$) ranked in the top 4 of both influence and affectedness.

Regarding the importance of risk-influencing factors, the outcome factors result from a combination of other factors that directly impact corporate financial risk in the digital context. The causal factors are mainly inadequate talent training and incentive mechanism ($x_{22}$), the accounting department is co-located with the rest of the departments ($x_{25}$), and small scale enterprise operation ($x_9$). The result factors mainly include financial data leakage, loss or falsification ($x_{26}$), the cash flow situation is not optimistic ($x_8$), the financial staff's mismanagement of digital products ($x_{24}$). Meanwhile, the centrality of these two influencing factors is also larger, which has a strong negative effect on enterprise financial risk management.

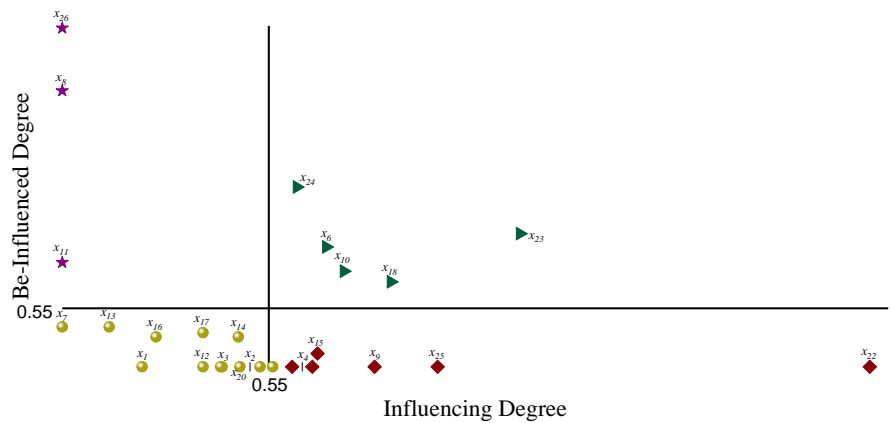

Figure 3 Influence-influenced diagram of financial risk in digital transformation

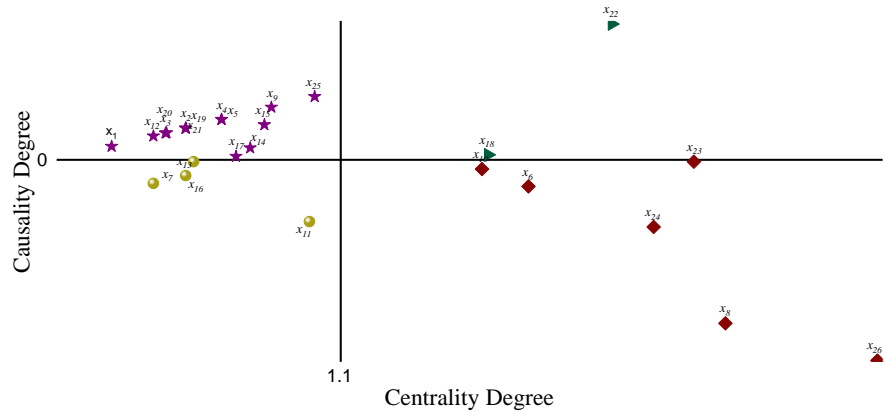

Figure 4 Centrality-causality diagram of financial risk in digital transformation

## 4.3 ISM Result

Based on DEMATEL, the ISM method is used to construct a multi-level recursive structure model to analyse further the influence logic and hierarchical relationship between the influencing factors. In the calculation process of ISM, the reachability matrix indicates whether there is a connection path between factors. If all the columns are 0, the factor only influences other factors and is not influenced by other factors, and vice versa if all the rows are 0. Based on this, the reachable set contains the factors that can affect other factors, i.e., the causal factors in the lower level. The prior set includes the factors that are influenced by other factors, i.e., the result factors in the upper level; and the intersection set represents the transition factors in the middle position. The results are shown in **Tab. 3**.

Table 3 Table of reachable sets and prior sets and their intersections

| Code | Reachable set | Antecedent set | Intersection |
|---|---|---|---|
| $x_1$ | 1,6,8,10,18 | 1 | 1 |
| $x_2$ | 2,11 | 2 | 2 |
| $x_3$ | 3,6,8,10,18 | 3 | 3 |
| $x_4$ | 4,6,8,10,18 | 4 | 4 |
| $x_5$ | 5,6,8,10,18 | 5 | 5 |
| $x_6$ | 6,8,10,18 | 1,3,4,5,6 | 6 |
| $x_7$ | 7 | 7,12 | 7 |
| $x_8$ | 8 | 1,3,4,5,6,8,10,13,14,15,17,18,19,20,22 | 8 |
| $x_9$ | 9,23,24,26 | 9 | 9 |
| $x_{10}$ | 8,10 | 1,3,4,5,6,10,18 | 10 |
| $x_{11}$ | 11 | 2,11,14,15,16,22 | 11 |
| $x_{12}$ | 7,12 | 12 | 12 |
| $x_{13}$ | 8,13 | 13,20 | 13 |
| $x_{14}$ | 8,11,14,17 | 14,15,22 | 14 |
| $x_{15}$ | 8,11,14,15,16,17 | 15,22 | 15 |
| $x_{16}$ | 11,16 | 15,16,22 | 16 |
| $x_{17}$ | 8,17 | 14,15,17,22 | 17 |
| $x_{18}$ | 8,10,18 | 1,3,4,5,6,18 | 18 |
| $x_{19}$ | 8,19 | 19 | 19 |
| $x_{20}$ | 8,13,20 | 20 | 20 |
| $x_{21}$ | 21,26 | 21 | 21 |
| $x_{22}$ | 8,11,14,15,16,17,22,23,24,26 | 22 | 22 |
| $x_{23}$ | 23,24,26 | 9,22,23 | 23 |
| $x_{24}$ | 24,26 | 9,22,23,24 | 24 |
| $x_{25}$ | 25,26 | 25 | 25 |
| $x_{26}$ | 26 | 9,21,22,23,24,25,26 | 26 |

As seen from **Fig. 5**, the factors influencing the financial risk of enterprises in the digital context are divided into five layers: surface factor L1, transition factors L2~4, and bottom factor L5. Surface factors play a direct role in influencing the financial risk of enterprises, while financial data leakage, loss or falsification ($x_{26}$), and unpromising cash flow situation ($x_8$) have the highest degree of influence and centrality, indicating that these two factors directly influence enterprise financial risk more than the other two factors. The underlying factors are the fundamental influencing factors for the emergence of corporate financial distress. Poor talent training and incentive mechanisms ($x_{22}$) has the largest centrality to the lack of enterprise technology innovation ability ($x_{15}$) has an impact, indicating that the talent cultivation dilemma is the most fundamental obstacle to enterprise financial management. Among the intermediate transition factors, the lack of digitally literate talents ($x_{23}$) and financial personnel's mismanagement of digital products ($x_{24}$) are the two factors with the highest centrality and being influenced.

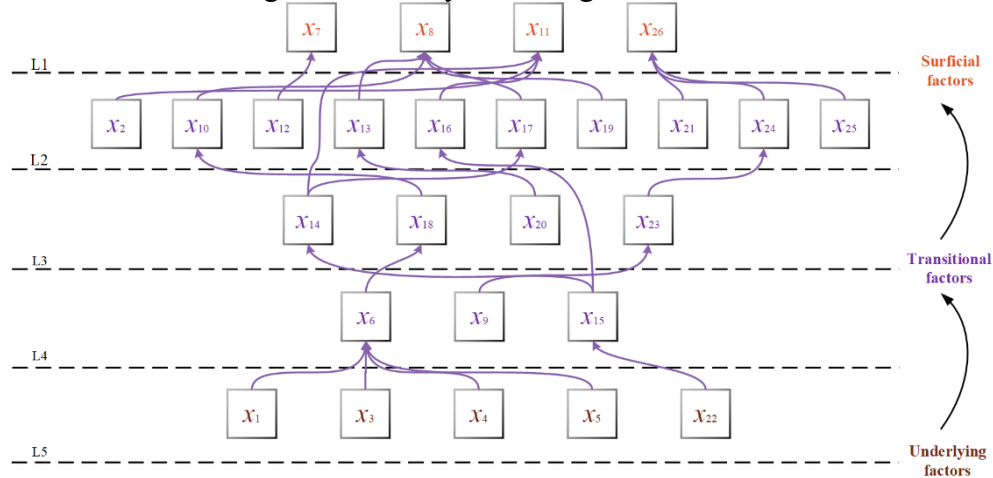

Figure 5 Hierarchy of the factors influencing the financial risk in digital transformation

## 4.4 MICMAC Result

The MICMAC model is constructed to deeply delineate the position and role of the factors influencing the financial risk of enterprises in the digital context and analyse the driving force and dependency of the factors. The results are shown in **Tab. 4**.

Table 4 Driving power and Dependence power

| Code | Indicator name | Driving | Dependence |
|---|---|---|---|
| $x_1$ | Unreasonable economic policy guidance | 5 | 1 |
| $x_2$ | Slow implementation of government subsidies and incentives | 2 | 1 |
| $x_3$ | Regulations related to digital development are unclear | 5 | 1 |
| $x_4$ | The business is located in an underdeveloped area | 5 | 1 |
| $x_5$ | The market management in the industry is chaotic | 5 | 1 |
| $x_6$ | overall digitalisation rate of the market is low | 4 | 5 |
| $x_7$ | Unreasonable structure of assets and liabilities | 1 | 2 |
| $x_8$ | The cash flow situation is not optimistic | 1 | 15 |
| $x_9$ | Small business scale | 4 | 1 |
| $x_{10}$ | There is a product inventory that is slow to sell | 2 | 7 |
| $x_{11}$ | Digital transformation costs are too high | 1 | 6 |
| $x_{12}$ | The investment planning plan is unreasonable | 2 | 1 |
| $x_{13}$ | Forecasting Mistakes When Quantifying Investing | 2 | 2 |
| $x_{14}$ | Low level of digitalisation of core business | 4 | 3 |
| $x_{15}$ | Insufficient technological innovation ability of enterprises | 6 | 2 |
| $x_{16}$ | Blind pursuit of advanced digital products | 2 | 3 |
| $x_{17}$ | Product market share reduction | 2 | 4 |
| $x_{18}$ | Inefficient cooperation with upstream and downstream enterprises | 3 | 6 |
| $x_{19}$ | Low credit leads to financing difficulties | 2 | 1 |
| $x_{20}$ | The enterprise transformation plan is not clear | 3 | 1 |
| $x_{21}$ | Internal information management system confusion | 2 | 1 |
| $x_{22}$ | Talent training and incentive mechanism is not perfect | 10 | 1 |
| $x_{23}$ | Lack of digitally literate talent | 3 | 3 |
| $x_{24}$ | Financial personnel make mistakes in operating digital products | 2 | 4 |
| $x_{25}$ | The accounting department is co-located with other departments | 2 | 1 |
| $x_{26}$ | Financial data leaked, lost, or falsified | 1 | 7 |

  The final factors will be divided into four categories in the form of two-dimensional axes located in 4 different quadrants, as shown in Fig. 6. The meanings of each quadrant are autonomous factors, dependent factors, and linkage and driver factors. Autonomous factors have weaker drivers and dependencies. Dependent factors have stronger dependencies and weaker drivers, and linkage factors have stronger drivers and dependencies. Drivers factor has stronger drivers and weaker dependencies.

  The overall market digitisation rate is less coordinated ($x_6$) is a linkage factor with a strong driving force and dependency, which is influenced by the external environment of the enterprise, but can affect the business ability of the enterprise. Inadequate talent training and incentive mechanisms ($x_{22}$), and insufficient enterprise technology innovation ability ($x_{15}$) are independent factors which have driving effects on other factors. Most of the autonomous factors come from the organisational structure factors of the enterprise, indicating that the human resources and management structure of the enterprise are transitional factors in this influence system. Among the dependency factors, the poor cash flow situation ($x_8$) is particularly important and most influenced by other factors and is also the top-level direct influence factor of financial risk in the ISM hierarchy.

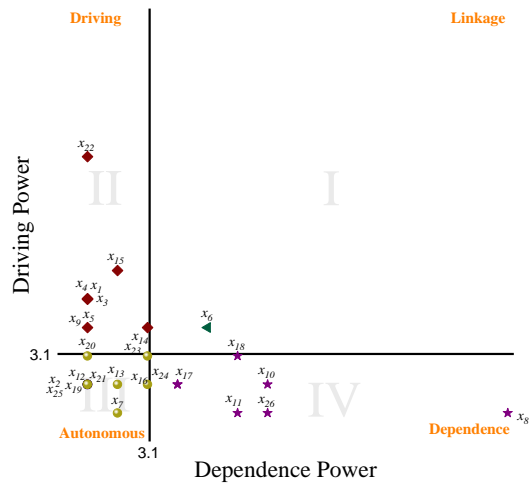

Figure 6 Driving-dependence diagram of financial risk in digital transformation

In summary, the system of financial risk factors in the digital context can be constructed as a five-level hierarchy divided into four categories: independent, linked, dependent, and autonomous. The role of risk factors can be reflected in the hierarchical path by combining the magnitude of each factor's centrality and influence degree. Among the bottom-level factors are imperfect talent training and incentive mechanism ($x_{22}$) has the highest influence degree, cause degree, centrality degree, and strongest driving force, and is the most fundamental obstacle factor. Among the surface factors are leaked, lost, or falsified financial data ($x_{26}$), and unpromising cash flow situation ($x_8$) are the two factors with the highest degree of being influenced, the highest degree of centrality, the lowest degree of cause, and the strongest dependence, and are the most direct risk influencing factors. Among the transition factors is the lack of digitally literate personnel ($x_{23}$), and financial personnel's failure to operate digital products ($x_{24}$) have greater centrality and higher influence degree and being influenced degree, and stronger driving forces among the autonomous factors, so they are essential to transition influencing factors.

# 5 Discussion

## 5.1 Importance of influencing factors

In the financial risk impact factor system constructed in this paper, the factors that have the strongest direct correlation with the risk are mainly the financial data leakage problem and poor cash flow situation, which indicates that the quality of financial data and the situation of liquid corporate assets are the important causes of the financial sewing risk of the enterprise. From the perspective of the relevant influence relationship, the lack of talent training and incentive mechanism strongly influences other factors, the lack of digitally literate talent is a more important influence factor, and many aspects can influence the phenomenon of financial data leakage, loss, or falsification. From the analysis of the enterprise's organisational structure, the neglect of staff training and talent introduction can easily cause the reduction of digital talent, which in turn leads to the impact on the enterprise's financial work and eventually to financial risks. Therefore, financial data management, liquidity crisis prevention, and digital talent development are pivotal in curbing financial risks (Karkošková, 2022).

## 5.2 Hierarchy of influencing factors

Through the structural analysis of the influencing factors of enterprise financial risk systems in the digital background, this paper finds that the influencing factors show a multi-level distribution. Overall, the most important native influencing factor of enterprise financial risk is the inadequate personnel training and incentive mechanism. It mainly leads to the lack of digitally literate personnel. The resulting staff operational errors will further lead to the leakage, loss or falsification of financial data and the unpromising cash flow situation, which directly affects the enterprise's financial risk.

From the perspective of the firm's external environment, political policies, regional and market factors are the underlying factors that influence the overall coordination rate of the market in which the firm is located to affect the firm's economic performance, which indicates that the firm's external environment is usually the underlying factor of the firm's financial risk. From the perspective of the financial status of the enterprise's assets and liabilities, the investment and financing behavior of the enterprise affected by the organisational effectiveness of the enterprise will affect the cash flow and asset and liability structure of the enterprise, which will directly lead to the financial distress of the enterprise, which indicates that the current financial status of the enterprise will directly affect the size of the financial risk of the enterprise. From the perspective of the firm's business capability, the firm's innovation capability affects the firm's core business and the efficiency of cooperation with suppliers, which in turn affects the firm's market share. From the enterprise's organisational structure perspective, the enterprise's inadequate employee training and reward and punishment mechanisms can trigger a digital talent gap (Cetindamar Kozanoglu & Abedin, 2021). In the context of digital transformation, employees who are not digitally literate can cause economic losses to the company due to operational risks.

## 5.3 Influencing factors dependence

As the most fundamental obstacle to enterprise financial risk, the imperfect talent training and incentive mechanism is the strongest driving influence factor. The lack of enterprise technology innovation ability will trigger the enterprise's operation and organization to manage risk. At the same time, this paper finds that most of the autonomy factors come from the organisational structure factors of enterprises, which is mainly due to the extended transmission mechanism of financial risks brought about by the talent training mechanism of enterprises. However, it also shows that this moderate risk transmission is easier to be ignored by enterprises and the market and should gradually attract the attention of enterprises in the context of intensified digital transformation (Gallardo-Gallardo et al., 2020).

Transition factors play a moderating role in the entire risk transmission chain. Therefore, their analysis cannot be neglected. The overall market digital rate coordination factor, which is both dependent and driven, is a noticeable transition factor for risk transmission, which is influenced by the external environment of the company on the one hand, and can affect the business ability of the company on the other. Combining the analysis results of driving force and dependency, among the transition factors in the ISM hierarchy, the lack of digitally literate personnel and financial personnel's failure to operate digital products have a higher position in the risk impact system. Meanwhile, the driving force is also stronger in the autonomy factor. This paper considers these two factors the most important transition impact factors enterprises should focus on when conducting risk avoidance.

## 5.4 Management Recommendations

Firstly, the public sector of governments worldwide should improve policies on the digital development of society and strive to promote the overall digital development of the market. The government public sector should improve the rule of law system concerning the digital development of the economic market, guide enterprises to make reasonable digital transformation behaviors and create a healthy and orderly development environment. Meanwhile, it is paid attention to the monopoly problem in the development of digital technology and avoid the vicious development of the industry as a whole. Improve laws, regulations, and regulatory mechanisms to guide market enterprises to transform in a lawful and compliant manner and promote the sustainable and healthy development of the digital economy.

Secondly, enterprises should gradually pay attention to developing a digital talent training strategy to improve the digital literacy of financial staff. According to the strategic development direction of the enterprise, the talent digital literacy system is established in the talent strategy of the enterprise. Only by establishing as soon as possible a coherent central axis business strategy model from enterprise development strategy to talent strategy and cultivating the organisational capabilities of inherited strategy and successive performance can enterprises promote the construction process of digital enterprise management through the high-quality development of digital talent literacy. Then, it can make enterprise operations realise cost reduction and efficiency.

Thirdly, enterprises focus on investment and financing management, adhere to a scientific and appropriate development strategy, and strengthen working capital management. Appropriate investment and financing strategies play a key role in maintaining the continuity of the enterprise's capital chain and ensuring the normal operation of the enterprise (Bhalla, 2008). The investment and financing strategy should be in line with the development stage of the company. The investment and financing strategy should be adapted to the development stage of the enterprise, and the enterprise should consider its financial situation and actual needs when acquiring digital technology and products and should not blindly follow the high-tech. Before investing and financing, enterprises should establish a sound investment and financing management system to minimise the potential risks caused by information asymmetry and misjudgment of market and industry factors.

Fourthly, enterprises gradually improve the ability to scientific and technological innovation, keep pace with the development of the digital era, and the market as a whole goes hand in hand. The digital transformation and upgrading of enterprises have become an irreversible trend. On the one hand, with the help of digital technology, enterprises can more accurately grasp the changes in market demand and expand their market share according to the accurate analysis of potential user needs. On the other hand, digital technology connects the originally scattered equipment, enterprises, and markets, not only to realise the linked development of internal R&D and sales and other aspects of enterprises but also to enhance the efficiency of cooperation between enterprises and expand the development space of the industry as a whole by strengthening the connection and interoperability between different enterprises and between enterprises and markets. Therefore, enterprises that fail to realise synchronous digital transformation with the market and society in time will face the risk of being eliminated.

## 6 Conclusion

In recent years, digital transformation has become an important engine for the rapid development of enterprises in various countries and brings new challenges to the financial management of enterprises. In this paper, firstly, through literature combing, the index system of influencing factors of enterprise financial risk in the context of digital transformation is constructed. Eight experts are invited to score the influencing relationships among factors. The importance of factors and the correlation between factors are analysed according to the DEMATEL model, and the hierarchical structural relationship of system factors is determined using the ISM model. Based on this method, the MICMAC model to mine the dependency and driving relationships among the influencing factors.

According to the research results, there are two main transmission paths of corporate financial management risks: first, the policy and market environment of enterprises are not optimistic, which leads to the restriction of their operation ability, thus directly leading to the deterioration of their asset-liability structure and cash flow; second, the talent training and promotion mechanism of enterprises is not perfect, which leads to the restriction of their innovation ability and the existence of digital talent gap, thus making it difficult for enterprises to adapt to the digital The transformation of the social development background, to the enterprise's financial management system to bring a blow, and ultimately lead to the financial risk of enterprises.

This study is the first to systematically list the influencing factors of financial risks in the context of enterprise digital transformation from four aspects: external environment, financial status, operational capability, and organisational structure of enterprises. At the same time, the DEMATEL-ISM-MICMAC hybrid approach is used to analyse the influencing factors of enterprises' financial risks, which effectively expands the library of methods for managing and studying the financial risks of enterprises in the context of digitalisation. The research in this paper provides a theoretical basis for how companies in various countries can cope with the challenges of digital transformation and also provides ideas for future academic research in digital finance.

However, this study still has certain limitations. The data in this paper are obtained from expert scoring. Although the authority and representativeness of the data are ensured as much as possible, there is still no way to avoid the subjective nature of expert opinions that make the results of the study deviate from the actual situation to some extent. Moreover, the method used in this paper is most applicable to stable systems (Jafari-Sadeghi et al., 2021). Therefore, future research can be based on the theory of fuzzy mathematics to solve the problem of uncertainty in the system; it can also calculate the enterprise financial risk index based on the influencing factors listed in this paper to construct a comprehensive evaluation system with multiple criteria (Gu & Wang, 2022).

**Author Contributions:** Conceptualisation, J.D.; methodology, J.D.; formal analysis, J.D.; data curation, J.D.; supervision, J.D.; writing -original draft preparation, J.D.; writing-review and editing, J.D. The authors have read and agreed to the published version of the manuscript. The authors have read and agreed to the published version of the manuscript.
**Method:** All methods were carried out by relevant guidelines and regulations.
**Funding:** This research received no external funding.
**Data Availability Statement:** The data for this study comes from a survey questionnaire of eight anonymous experts.
**Conflicts of Interest:** The authors declare that there are no conflicts of interest.

# References:

[1] Ahmad, N., & Qahmash, A. (2021). Smartism: Implementation and assessment of interpretive structural modeling. *Sustainability*, 13(16), 8801.
[2] Al-Blooshi, L., & Nobanee, H. (2020). Applications of artificial intelligence in financial management decisions: A mini-review. *Available at SSRN 3540140*
[3] Belás, J., Dvorský, J., Kubálek, J., & Smrčka, L. (2018). Important factors of financial risk in the SME segment. *Journal of International Studies*
[4] Berman, S. J. (2012). Digital transformation: opportunities to create new business models. *Strategy & leadership*
[5] Bhalla, V. K. (2008). *Investment Management (Security Analysis and Portfolio Management)*. S. Chand Publishing.
[6] Cao, D., & Zen, M. (2005). An empirical analysis of factors influencing financial risk of listed companies in China. *Techno Economics & Management Research*, 6, 37-38.
[7] Cetindamar Kozanoglu, D., & Abedin, B. (2021). Understanding the role of employees in digital transformation: conceptualization of digital literacy of employees as a multi-dimensional organizational affordance. *Journal of Enterprise Information Management*, 34(6), 1649-1672.
[8] Chen, J. (2021). Improved DEMATEL-ISM integration approach for complex systems.

*Plos One*, 16(7), e254694.

[9] Chollet, P., & Sandwidi, B. W. (2018). CSR engagement and financial risk: A virtuous circle? International evidence. *Global Finance Journal*, 38, 65-81.

[10] Correani, A., De Massis, A., Frattini, F., Petruzzelli, A. M., & Natalicchio, A. (2020). Implementing a digital strategy: Learning from the experience of three digital transformation projects. *California Management Review*, 62(4), 37-56.

[11] Damerji, H., & Salimi, A. (2021). Mediating effect of use perceptions on technology readiness and adoption of artificial intelligence in accounting. *Accounting Education*, 30(2), 107-130.

[12] Dai, Y., & Lu, Z. (2023). Regional digital finance and corporate financial risk: based on Chinese listed companies. Emerging Markets Finance and Trade, 59(2), 296-311.

[13] Du, Y., & Li, X. (2021). Hierarchical DEMATEL method for complex systems. Expert Systems with Applications, 167, 113871.

[14] Fu, G., Fu, W., & Liu, D. (2012). Empirical study on financial risk factors: Capital structure, operation ability, profitability, and solvency——evidence from listed companies in China. *J. Bus. Manag. Econ*, 3, 173-178.

[15] Gallardo-Gallardo, E., Thunnissen, M., & Scullion, H. (2020). Talent management: context matters (31, pp. 457-473): Taylor & Francis.

[16] Goerzig, D., & Bauernhansl, T. (2018). Enterprise architectures for the digital transformation in small and medium-sized enterprises. *Procedia Cirp*, 67, 540-545.

[17] Gonçalves, M. J. A., Da Silva, A. C. F., & Ferreira, C. G. (2022*The future of accounting: how will digital transformation impact the sector?* Paper presented at the Informatics.

[18] Gu, W., & Wang, J. (2022). Research on index construction of sustainable entrepreneurship and its impact on economic growth. *Journal of Business Research*, 142, 266-276.

[19] Hacioglu, U., & Aksoy, T. (2021). *Financial Ecosystem and Strategy in the Digital Era: Global Approaches and New Opportunities*. Springer.

[20] Ionaşcu, I., Ionaşcu, M., Nechita, E., Săcărin, M., & Minu, M. (2022). Digital transformation, financial performance and sustainability: Evidence for European Union listed companies. *Amfiteatru Economic*, 24(59), 94-109.

[21] Jafari-Sadeghi, V., Mahdiraji, H. A., Bresciani, S., & Pellicelli, A. C. (2021). Context-specific micro-foundations and successful SME internationalisation in emerging markets: A mixed-method analysis of managerial resources and dynamic capabilities. *Journal of Business Research*, 134, 352-364.

[22] Karkošková, S. (2022). Data governance model to enhance data quality in financial institutions. *Information Systems Management*, 1-21.

[23] Kraus, S., Jones, P., Kailer, N., Weinmann, A., Chaparro-Banegas, N., & Roig-Tierno, N. (2021). Digital transformation: An overview of the current state of the art of research. *Sage Open*, 11(3), 1915380888.

[24] Leeflang, P. S., Verhoef, P. C., Dahlström, P., & Freundt, T. (2014). Challenges and solutions for marketing in a digital era. *European Management Journal*, 32(1), 1-12.

[25] Luo, Y., Peng, Y., & Zeng, L. (2021). Digital financial capability and entrepreneurial performance. *International Review of Economics & Finance*, 76, 55-74.

[26] Mahtani, U. S., & Garg, C. P. (2018). An analysis of key factors of financial distress in airline companies in India using fuzzy AHP framework. *Transportation Research Part A: Policy and Practice*, 117, 87-102.

[27] Mani, V., Agrawal, R., & Sharma, V. (2016). Impediments to social sustainability adoption in the supply chain: An ISM and MICMAC analysis in Indian manufacturing industries. *Global Journal of Flexible Systems Management*, 17, 135-156.

[28] Morgan, P. J., Huang, B., & Trinh, L. Q. (2019). The need to promote digital financial literacy for the digital age. *IN THE DIGITAL AGE*

[29] Mosteanu, N. R., & Faccia, A. (2020). Digital systems and new challenges of financial management–FinTech, XBRL, blockchain and cryptocurrencies. *Quality-Access to Success Journal*, 21(174), 159-166.

[30] Sindhwani, R., & Malhotra, V. (2017). Modelling and analysis of agile manufacturing system by ISM and MICMAC analysis. *International Journal of System Assurance Engineering and Management*, 8, 253-263.

[31] Suling, F., Shu, Z., & Haoyue, W. (2021). The Impact of Fintech on Corporate Financial Risk and Its Internal Mechanism: On the Threshold Effect of Financial Regulation. *Reform*(10), 84-100.

[32] Tabrizi, B., Lam, E., Girard, K., & Irvin, V. (2019). Digital transformation is not about technology. *Harvard Business Review*, 13(March), 1-6.

[33] Teichert, R. (2019). Digital transformation maturity: A systematic review of literature. *Acta universitatis agriculturae et silviculturae mendelianae brunensis*

[34] Vovchenko, N. G., Andreeva, O. V., Orobinskiy, A. S., & Sichev, R. A. (2019). Risk control in modeling financial management systems of large corporations in the digital economy

[35] Westerman, G., Tannou, M., Bonnet, D., Ferraris, P., & McAfee, A. (2012). The Digital Advantage: How digital leaders outperform their peers in every industry. *MITSloan Management and Capgemini Consulting, MA*, 2, 2-23.

[36] Wu, W., & Lee, Y. (2007). Developing global managers' competencies using the fuzzy DEMATEL method. *Expert Systems with Applications*, 32(2), 499-507.

[37] Zeng, D., Cai, J., & Ouyang, T. (2021). Research on digital transformation: integration framework and future prospects. *Foreign Econ Manag*, 5, 63-76.

[38] Zhang, T., Shi, Z., Shi, Y., & Chen, N. (2022). Enterprise digital transformation and production efficiency: Mechanism analysis and empirical research. *Economic Research-Ekonomska Istraživanja*, 35(1), 2781-2792.

[39] Zhao, N., Song, Z., Li, P., & Shi, Z. (2022). The Influence of Digital Transformation on Enterprise Financial Risk. Scientific Decision Making(12), 21-36.